\documentclass[pdflatex,sn-mathphys-num]{sn-jnl}

\usepackage{graphicx}%
\usepackage{multirow}%
\usepackage{amsmath,amssymb,amsfonts}%
\usepackage{amsthm}%
\usepackage{mathrsfs}%
\usepackage[title]{appendix}%
\usepackage{xcolor}%
\usepackage{textcomp}%
\usepackage{manyfoot}%
\usepackage{booktabs}%
\usepackage{algorithm}%
\usepackage{algorithmicx}%
\usepackage{algpseudocode}%
\usepackage{listings}%
\usepackage{subcaption}
\usepackage{lineno}

\theoremstyle{thmstyleone}%
\theoremstyle{thmstyletwo}%

\theoremstyle{thmstylethree}%

\begin{document}

\title[Article Title]{Far-field excitation of symmetry-protected bound states in the
continuum by nonlinear virtual sources}


\author[1,2]{\fnm{Marc} \sur{Martí-Sabaté}}
\equalcont{These authors contributed equally to this work.}

\author[3]{\fnm{Yijie} \sur{Zhang}}
\equalcont{These authors contributed equally to this work.}

\author[4,5]{\fnm{Shengming} \sur{Sun}}

\author[3]{\fnm{Ruxin} \sur{Li}}

\author[5]{\fnm{Yabin} \sur{Jin}}

\author*[3]{\fnm{Junfei} \sur{Li}}\email{junfeili@purdue.edu}

\author*[4]{\fnm{Daniel} \sur{Torrent Martí}}\email{dtorrent@uji.es}

\affil[1]{Department of Mathematics, Imperial College London, London SW7 2AZ, UK}

\affil[2]{Instituto Universitario de Matem\'atica  Pura y Aplicada, 
Universitat Polit\`ecnica de Val\`encia, 46022 (Spain)}

\affil*[3]{\orgdiv{School of Mechanical Engineering}, \orgname{Purdue University}, \orgaddress{\street{177 S Russell St. HLAB 1020}, \city{West Lafayette}, \postcode{47907}, \state{Indiana}, \country{USA}}}

\affil*[4]{\orgdiv{GROC, UJI, Institut de Noves Tecnologies de la Imatge (INIT)}, \orgname{Universitat Jaume I}, \orgaddress{\city{Castelló}, \postcode{12071}, \country{Spain}}}

\affil*[5]{\orgdiv{Institute of Computational Mechanics x AI \& College of Intelligent Robotics and Advanced Manufacturing}, \orgname{Fudan University}, \orgaddress{ \city{Shanghai}, \postcode{200433}, \state{Shanghai}, \country{China}}}



\abstract{Bound states in the continuum (BICs) enable complete wave confinement within open systems through destructive interference or symmetry protection, giving rise to ideally infinite quality factors and extreme field enhancement. Their practical exploitation, however, is hindered by a fundamental paradox: the same mechanism that suppresses radiation also prevents efficient excitation from the far field. Existing experimental approaches typically overcome this limitation by introducing external coupling channels that inevitably perturb the protected state and reduce its confinement.

Here we demonstrate the non-invasive excitation of a symmetry-protected acoustic BIC through nonlinear virtual sources generated directly at the surface of the structure. By focusing time-modulated ultrasonic beams onto a metamirror, nonlinear demodulation produces localized low-frequency sources whose spatial phase matches the symmetry of the target mode while avoiding any physical coupling to the resonator. This approach injects energy into the BIC without opening additional radiative channels, preserving its intrinsic non-radiative character. We develop a theoretical framework describing the formation and symmetry protection of the mode, realize a compact experimental implementation through symmetry reduction, and directly measure the confined pressure field and its quality factor. Our results establish nonlinear virtual sources as a route to accessing symmetry-protected states in open systems while maintaining their exceptional confinement properties, providing a general strategy for wave manipulation and high-sensitivity resonant devices across physical platforms.}




\maketitle

\section{Introduction}\label{sec1}

Controlling the propagation, confinement, and localization of waves in open systems is a central challenge across many areas of physics and engineering. The ability to concentrate energy in space while suppressing radiation losses impacts a broad range of technologies, including sensing, imaging, communications, and information processing. In resonant systems, strong confinement is typically accompanied by enhanced interactions between waves and matter, leading to improved device performance. However, achieving long-lived resonances in structures that remain accessible to external excitation and measurement is fundamentally difficult because coupling to the surrounding environment inevitably introduces radiation losses.

Bound states in the continuum (BICs) provide a remarkable solution to this problem. First predicted by von Neumann and Wigner in the context of quantum mechanics \cite{von1993merkwurdige}, BICs are localized eigenstates whose frequencies lie within the radiation continuum while remaining perfectly confined. Despite coexisting with propagating waves, these states do not radiate energy due to destructive interference, symmetry incompatibility, or topological protection mechanisms. During the last decade, BICs have emerged as a unifying concept across wave physics and have been extensively investigated in photonic \cite{molina2012surface, miroshnichenko2010fano, hsu2016bound, hsu2013bloch, sadreev2021interference, bulgakov2008bound, zhen2014topological}, acoustic \cite{huang2022general, quotane2018trapped, jin2017tunable, mizuno2019fano, sadreev2022degenerate, marti2024observation}, and elastic systems \cite{pagneux2013trapped, cao2025asymmetric, an2024multibranch, marti2023bound, wen2025compact, gao2024bound}. Their ideally infinite lifetime and divergent quality factor make them attractive for applications requiring extreme field enhancement, narrow spectral selectivity, and enhanced wave–matter interactions \cite{kodigala2017lasing, koshelev2019meta, wu2020room, koshelev2020subwavelength}.

Among the different classes of BICs, symmetry-protected (SP) BICs are particularly appealing because their existence is guaranteed by symmetry considerations rather than fine parameter tuning \cite{li2019symmetry, sadrieva2019experimental, cong2019symmetry}. In these systems, the mode symmetry is incompatible with the symmetry of the available radiation channels, preventing energy leakage into the far field. However, this same protection mechanism introduces a fundamental experimental challenge: an ideal SP-BIC cannot be excited through conventional linear coupling. Consequently, most experimental realizations rely on introducing intentional perturbations, such as structural asymmetries or external waveguide couplings, which transform the ideal BIC into a quasi-BIC with finite radiative losses \cite{huang2022general,jia2023bound,marti2024observation}. Although these approaches enable indirect characterization through transmission or reflection measurements, they necessarily modify the intrinsic properties of the state and limit direct access to its field distribution.

In acoustics, experimental investigations of BICs have predominantly focused on cavity-waveguide systems and partially open resonators, where the existence of the bound state is inferred from scattering measurements \cite{huang2021sound,huang2022general,jia2023bound}. More recently, direct visualization of acoustic quasi-BIC fields has been achieved using transparent resonators and laser Doppler vibrometry \cite{kronowetter2023realistic}. Despite these advances, experimental demonstrations remain largely restricted to geometries where external coupling channels are deliberately introduced. As a result, direct access to the fields of an ideal symmetry-protected BIC in a fully open acoustic structure remains an outstanding challenge. Addressing this limitation is essential for exploiting the exceptional field confinement and enhancement associated with BICs in practical sensing, imaging, and nonlinear-wave applications.

An alternative route to access otherwise inaccessible bound states is provided by nonlinear interactions \cite{chukhrov2021excitation,calajo2019exciting}. Unlike linear excitation mechanisms, nonlinear processes can create additional frequency channels whose symmetry properties differ from those of the incident field. Recent theoretical studies have suggested that second-harmonic generation and other wave-mixing processes can selectively populate BICs without introducing direct radiative coupling \cite{yuan2020excitation}. Moreover, the growing interest in Kerr-effect nonlinearities \cite{krasikov2018nonlinear,chukhrov2021excitation}, optomechanical \cite{yu2022observation}, and multiphoton scattering \cite{calajo2019exciting} highlights the broader potential of exploiting nonlinear physics to control high-Q states. Nevertheless, the experimental realization of nonlinear excitation of a symmetry-protected BIC has remained elusive, particularly in acoustic metasurfaces operating in open environments.

Here, we demonstrate the design, realization, and experimental characterization of a symmetry-protected BIC supported by a fully open acoustic metasurface. By employing a time-modulated ultrasonic excitation scheme that generates the resonant mode through second-harmonic generation, we access the BIC without introducing an external coupling channel that would compromise its symmetry protection. This nonlinear excitation mechanism enables direct measurement of the acoustic field associated with the bound state and reveals its robustness against fabrication imperfections. Our results establish nonlinear wave conversion as an effective strategy for accessing otherwise inaccessible bound states and open new opportunities for the implementation of BIC-enhanced acoustic devices for sensing, imaging, and wave manipulation.

\section{Results}\label{sec2}

\begin{figure}[h!]
    \makebox[\textwidth][c]{\includegraphics[width=5.5in,height=3.4in]{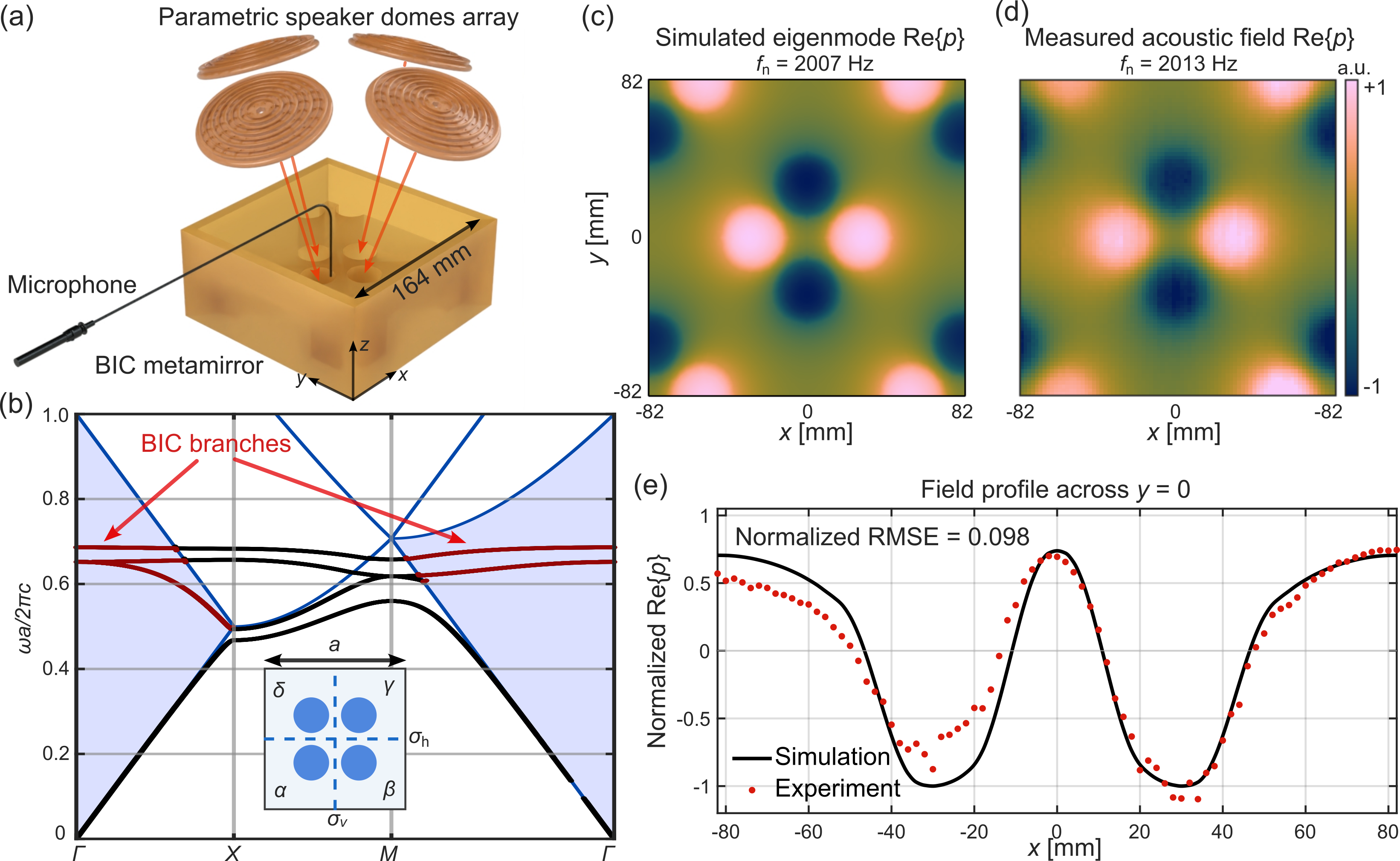}}
    \caption{\textbf{SP BIC experimental measurement.} \textbf{a.} Schematic of the designed device supporting SP-BICs. \textbf{b.} Band structure for the periodic crystal with BIC branches indicated in red. \textbf{c.} FEM simulated eigenmode. \textbf{d.} Experimentally measured BIC. \textbf{e.} Field profile comparison at $y = 0$ between simulation and experiment.}
    \label{fig:figure_1_main_article}
\end{figure}

A symmetry-protected BIC is engineered in 
Fig. \ref{fig:figure_1_main_article}. We consider a rigid surface (such that no elastic waves are excited within it) patterned with an infinite periodic array of unit cells, each containing four identical resonators (boreholes) arranged in a square configuration (Fig. \ref{fig:figure_1_main_article}(b) inset). The acoustic pressure field in the bulk is expanded in reciprocal Bloch modes of the lattice. Within each resonator, only the fundamental resonance is retained, as higher-order modes lie well outside the frequency range of interest. Using a mode-matching approach \cite{torrent2012acoustic,torrent2018acoustic,marti2024observation}, and enforcing continuity of both pressure and normal velocity field at the interface, we obtain a secular equation whose numerical solution yields the band structure shown in Fig. \ref{fig:figure_1_main_article}(b) (see Appendix \ref{secA1} for details).

In Fig. \ref{fig:figure_1_main_article}(b), both the dispersion bands and the sound cone are shown. The latter (dashed blue lines) marks the onset of propagating modes in the bulk, where the in-plane wavevector is proportional to frequency through the speed of sound. Below the sound cone, the bands (black lines) correspond to surface acoustic waves that remain confined to the structure and do not radiate into the bulk. Above the sound cone, the bands lie within the radiation continuum and correspond to leaky surface modes that couple to propagating bulk waves.  

The BIC originates from a band-folding mechanism. Consider first a configuration in which the resonators are located at $(\pm a/4, \pm a/4)$. In this case, the chosen unit cell is a supercell, rather than the primitive cell of the lattice. The corresponding Brillouin zone is therefore reduced, leading to band folding. In the unfolded (primitive) description, the $X$ point would lie at at $2\pi/a$ rather than $\pi/a$, and the modes appearing at the $\Gamma$ point and in Fig. \ref{fig:figure_1_main_article}(b) would instead reside at the $X$ point. These modes would then correspond to propagating states below the sound cone.

We now introduce a perturbation by displacing the resonators to positions $(\pm (a/4 + \varepsilon), \pm (a/4 +\varepsilon))$, with $|\varepsilon| \ll a/4$, while preserving the $C_4$ symmetry of the unit cell, as shown in Fig. \ref{fig:figure_1_main_article}(b) inset. This perturbation breaks the supercell description, so that the four-hole configuration becomes the primitive unit cell. As a result, the folded bands acquire physical significance and appear above the sound cone, within the radiation continuum, while retaining the propagation characteristics inherited from the supercell description.

The $C_4$ rotational symmetry allows the solution to be decomposed into discrete Fourier components of the form

\begin{equation}
    B_n = B_0e^{i2\pi n\ell/N}, \quad n,\ell \in \{0,1,2,3\},
    \label{eq:ansatz_main}
\end{equation}
where $N = 4$ is the number of resonators, $B_n$ denotes the amplitude of the fundamental resonance in each resonator, and $\ell$ is the azimuthal mode index characterizing the phase symmetry of the collective mode. In the first radiation zone, the scattered field consists of evanescent components in the out-of-plane direction, except for the fundamental diffraction order ($\mathbf{G} = \mathbf{0}$). The existence of a BIC therefore requires the cancellation of this radiative channel, that is, vanishing of the coefficient $A_0$.

For normal incidence ($\mathbf{K} = \mathbf{0}$), this coefficient can be written as

\begin{equation}
    A_0 = f\sin{(k_0L)}\sum_{\alpha=0}^{3}B_\alpha,
    \label{eq:radiation_cancellation_condition}
\end{equation}
where $f$ is the filling fraction of the resonators, $k_0 = \omega/c_b$ is the bulk wavenumber, and $L$ is the resonator's depth. Substituting the ansatz in equation (\ref{eq:ansatz_main}) shows that, for all modes with $\ell \neq 0$, the sum vanishes due to destructive interference. As a result, monopolar radiation into the far field is suppressed, and a symmetry-protected BIC is formed (see Appendix \ref{secA1} for derivations). 

The above discussion assumes normal incidence. More generally, symmetry-protected BICs persist for oblique incidence ($\mathbf{K} \neq \mathbf{0}$) provided that the incident wavevector preserves a mirror symmetry of the unit cell. In Fig. \ref{fig:figure_1_main_article}(b), this condition is satisfied along the $\Gamma-X$ direction ($\mathbf{K} = K\mathbf{\hat{x}}$) and the $M-\Gamma$ direction ($\mathbf{K} = K(\mathbf{\hat{x}}+\mathbf{\hat{y}})$), which retain symmetry with respect to orthogonal axes. Along these directions, bands within the radiation continuum that satisfy the phase-mismatch condition at $\Gamma$ remain non-radiative. These states propagate along the surface while remaining decoupled from bulk modes (see Appendix \ref{secA1} for a detailed analysis). 

Having established the theoretical framework, we next design the acoustic crystal to support a symmetry-protected BIC at low frequencies. The geometric parameters are chosen such that the target resonance occurs at approximately $1.8$ kHz. The resulting structure has a lattice constant $a = 12$ cm, resonator's depth $L = 3.45$ cm, resonator's radius $R = 1.5$ cm, and a resonator's displacement from the unit-cell centre along the $\hat{x}$ direction of $d = 2$ cm.

The mode profile shown in Fig. \ref{fig:figure_1_main_article}(c) exhibits mirror symmetry with respect to the $\pi/4$ and $3\pi/4$ directions. This observation enables a substantial reduction of the system size through the method of images \cite{jackson1999classical}. Specifically, the unit cell is rotated by $\pi/4$, and rigid boundaries (Neumann boundary conditions) are introduced along the edges of the resulting domain. These boundaries enforce mirror symmetry of the acoustic field and therefore preserve the symmetry class of the BIC mode. In this way, the infinite periodic crystal can be replaced by a finite structure while retaining the modes of interest. The resulting design is shown in Fig. \ref{fig:figure_1_main_article}(a). This structure constitutes a compact realization of the infinite crystal. Exploiting the symmetry of the BIC reduces the required experimental footprint from the meter scale expected for a conventional finite approximation of the periodic system to only $18\times 18~\mathrm{cm}^2$, while preserving the relevant modal characteristics.

We verify the validity of this reduced configuration using finite-element simulations performed in COMSOL Multiphysics. The calculated eigenmode spectrum confirms the existence of the symmetry-protected BIC. The corresponding eigenfrequency and pressure distribution are shown in Fig.\ref{fig:figure_1_main_article}(c).

\section{Experimental demonstration}\label{sec:experiment}

Experimental observation of a BIC requires overcoming its intrinsic decoupling from the radiation continuum. Consequently, direct excitation by an incident plane wave is forbidden in the ideal lossless limit and remains highly inefficient in realistic structures. To address this challenge, we employ a non-invasive excitation scheme based on nonlinear ultrasonic demodulation, which generates deeply subwavelength, low-frequency virtual sources directly at the metamirror surface. Unlike our previous implementation \cite{marti2024observation}, where the excitation was achieved through direct acoustic injection into the cavity, the present approach does not introduce additional radiation channels and therefore preserves the intrinsic confinement of the BIC.

The physical basis of this scheme is symmetry-matched excitation. The target BIC exhibits an antisymmetric spatial phase pattern across the four central holes, with adjacent holes oscillating out of phase. We therefore apply alternating phases to the audio-frequency modulation signals driving the four ultrasonic arrays. Through nonlinear demodulation, these inputs generate localized audio-frequency virtual sources that inherit the imposed phase pattern. The resulting four virtual sources reproduce the spatial symmetry of the target BIC, thereby enabling coupling to the otherwise inaccessible mode without opening an additional radiation channel.

\begin{figure}[h!]
    \makebox[\textwidth][c]{\includegraphics[width=5.5in,height=3.44in]{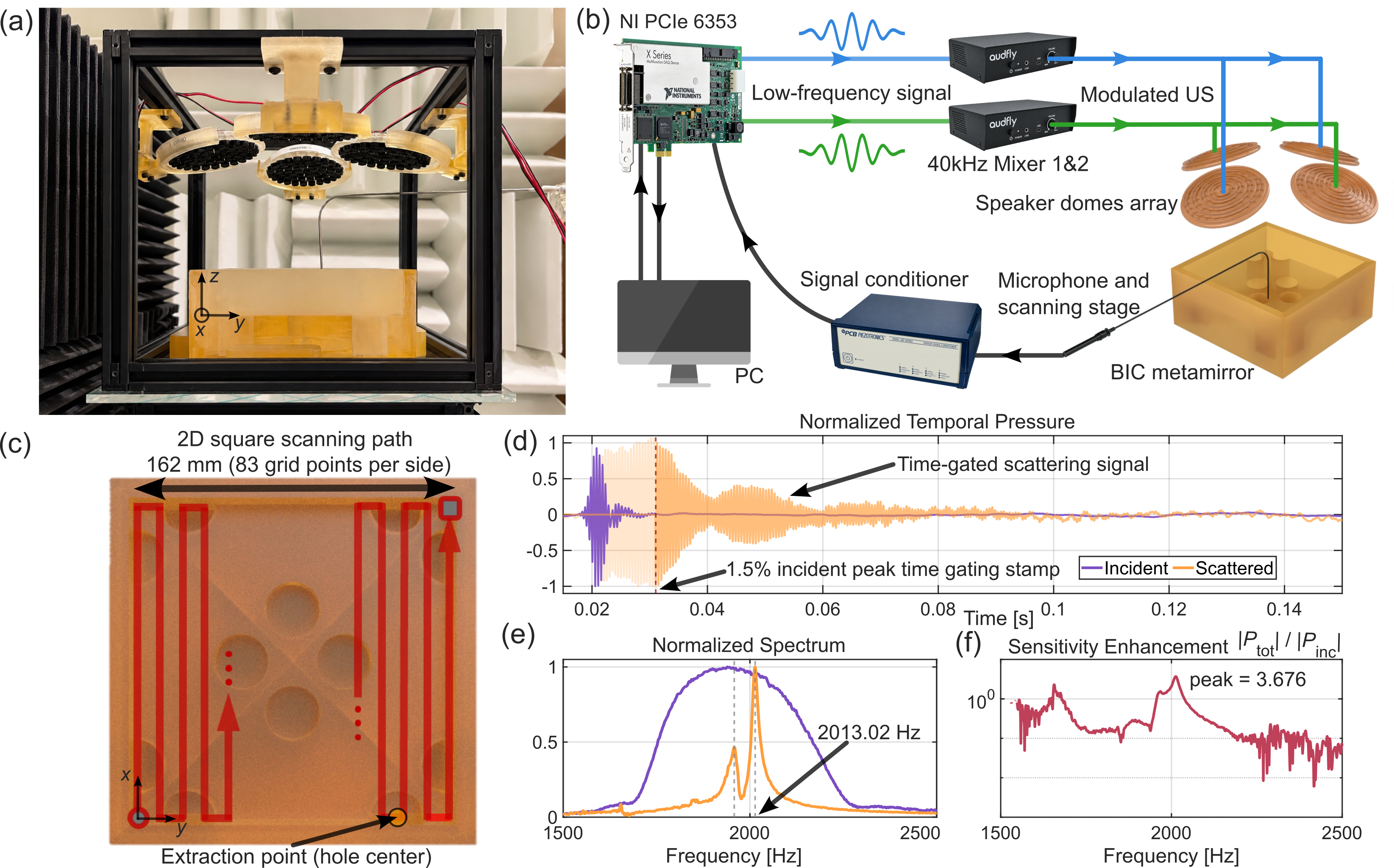}}
    \caption{\textbf{Experimental setup and spectral measurements.} \textbf{a.} Picture of the experimental setup with the excitation and measurement system. \textbf{b.} Schematics of the setup. \textbf{c.} Zenithal plane of the designed crystal with raster scan indicated. \textbf{d.} Temporal signal at the extraction point. Purple and orange lines indicate incident and scattered fields, respectively. \textbf{e.} Normalized spectral signal at the extraction point. \textbf{f.} Sensitivity enhancement characterized by the ratio between the total and incident fields.}
    \label{fig:figure_2_main_article}
\end{figure}

Experimentally, four spherical ultrasonic arrays focus energy onto the central holes, enhancing nonlinear demodulation and localizing the virtual sources at the target apertures. The resulting acoustic field was spatially reconstructed, as depicted in Fig.\ref{fig:figure_2_main_article}(a). The theoretical basis for the generation of the demodulated audio-frequency field and the preservation of its imposed phase pattern is detailed in Appendix \ref{secA2}, while the design and acoustic characterization of the excitation system is presented in Appendix \ref{secA3}. 

The field was scanned with a labyrinth route as outlined in Figure \ref{fig:figure_2_main_article}(c). Appendix \ref{secA4}.2,3,4 provide more detailed descriptions of  pulse definition, signal pathway, and scanning route. The scanned result is illustrated in Figure \ref{fig:figure_1_main_article}(d), showing the real pressure field. After normalizing the amplitude, the measured field faithfully reproduces the simulated BIC profile. For a better comparison, Fig. \ref{fig:figure_1_main_article}(e) shows the field profile at $y = 0$ for the eigenfrequency simulation from panel c and the experimental scan from panel d. The field profile shows an overall agreement in the mode shape, with a mean relative error of 0.098. Differences in the profile might arise from intrinsic material loss, or from the fact that the experimental measurement is a scattering measurement, while the simulation is an eigenfrequency solution. In the case of the experiment, the incident and the scattering field are not separable straightforwardly. Thus, a small contribution coming from the incident field might be present in the scattering measurements.   

Since the BIC field is excited by four parametric speaker domes focused onto the central holes, hard time gating was applied to the measured signal to suppress the contribution of the incident field and isolate the scattered BIC response (Appendix \ref{secA4}.5). Figure~\ref{fig:figure_2_main_article}(d) shows the time gating applied to the measured signal. The orange trace denotes the scattered field after time gating, whereas the purple trace denotes the incident field. The vertical dashed line indicates the temporal threshold used for the gate. The normalized spectra in Fig.~\ref{fig:figure_2_main_article}(e) compare the incident field with the time-gated scattered field. The time-gated spectrum is evaluated at the point of maximum field amplitude, marked by the orange dot in Fig.~\ref{fig:figure_2_main_article}(c). Following time gating, the broad incident spectrum spanning approximately $1.7$--$2.25\ \text{kHz}$ is reduced to two dominant scattered-field peaks at $2.013\ \text{kHz}$ and $1.959\ \text{kHz}$.

The normalized scattered-field map at $2.013\ \text{kHz}$ is shown in Fig.~\ref{fig:figure_1_main_article}(d), where the measured field is normalized by the peak amplitude of the incident signal. The resulting field distribution clearly reproduces the designed BIC confinement, with enhanced amplitudes at the hole locations. The corresponding field enhancement is shown in Fig.~\ref{fig:figure_2_main_article}(f). However, the apparently modest enhancement should not be interpreted as a direct quantitative measure of the intrinsic BIC amplification, because the incident reference is produced by direct excitation, whereas the scattered signal is generated through nonlinear modulation. The two fields therefore have different effective source strengths and are not generated under equivalent excitation conditions. Consequently, the reported value represents the measured system-level enhancement rather than the intrinsic modal enhancement of the BIC.

\begin{figure}[h!]
    \makebox[\textwidth][c]{\includegraphics[width=5.5in,height=2.44in]{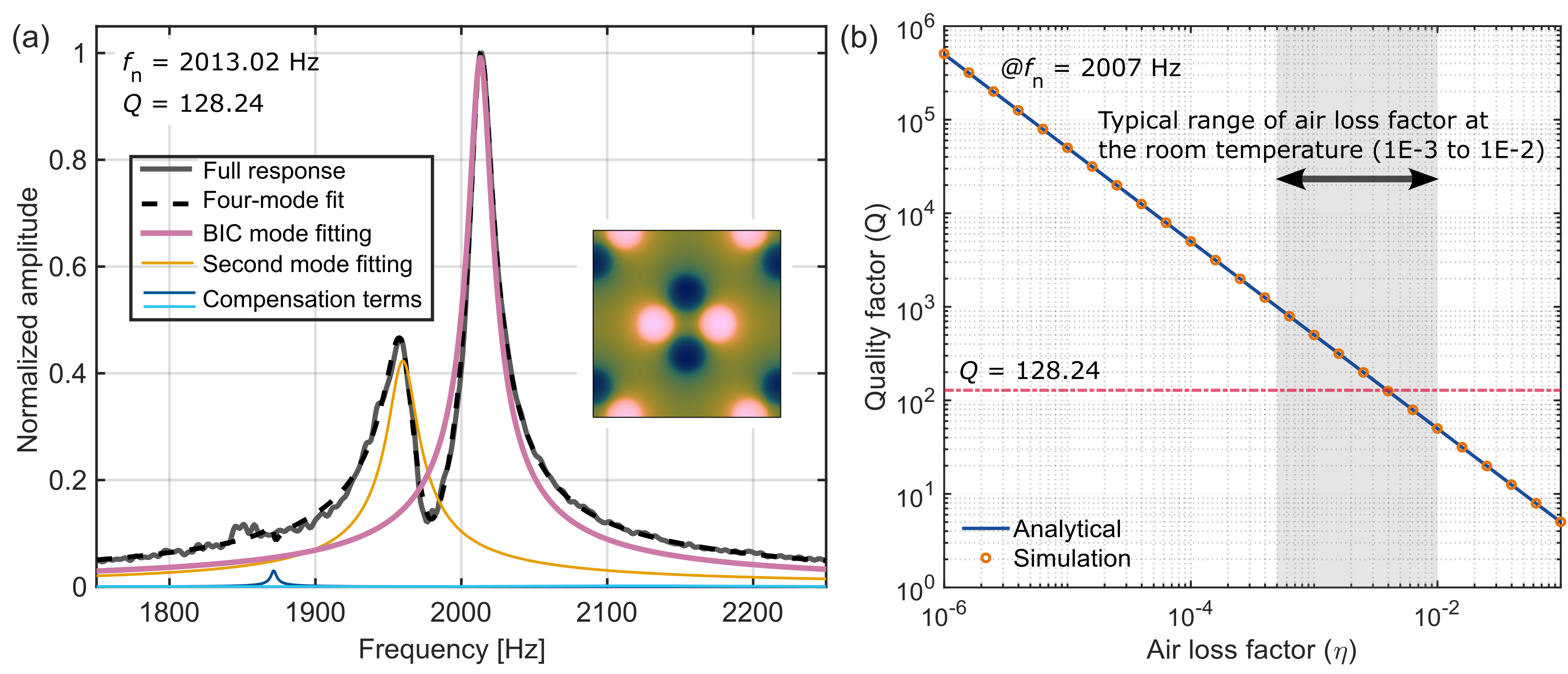}}
    \caption{$\mathbf{Q}$ \textbf{factor characterization.} \textbf{a.} Curve fitting for estimating $Q$ factor properties of the SP-BIC. \textbf{b.} Comparison of the experimental $Q$ factor with lossy FEM simulations.}
    \label{fig:figure_3_main_article}
\end{figure}

The quality factor (Q-factor) of the BIC is estimated using a curve-fitting procedure in the frequency domain, following a superposition of four Lorentzian shapes (Appendix \ref{secA4}.6) in the form of

\begin{equation}
    S(\omega) = \frac{B\exp(j\phi)}{j(\omega-\omega_r)+\omega_i}.
\end{equation}
The measured resonance frequency differs from the 
simulated value by $0.3\%$ ($6$ Hz), likely because of fabrication tolerances and the finite stiffness of the 3D-printed resin, which deviates from the perfectly rigid material assumed in the simulation. The extracted quality factor is independent of the measurement position.

The experimental estimation of the $Q$ factor can be decomposed in two contributions: the leakage to the continuum or radiative losses $Q_{rad}$ and the intrinsic losses (viscous, thermal) $Q_{loss}$. 

\begin{equation}
    \frac{1}{Q_{tot}} = \frac{1}{Q_{rad}} + \frac{1}{Q_{loss}}.    
\end{equation}

In this case, we are not opening a radiative channel, as we are exciting our mode from the far-field using a nonlinear interaction. Thus, $Q_{rad} \rightarrow \infty$. This leads us to $Q_{tot} = Q_{loss}$. With this in mind, our experimental guess let us estimate the amount of intrinsic loss in our system. Fig. \ref{fig:figure_3_main_article}(b) shows the Q factor of our eigenfrequency simulation in COMSOL as a function of the material losses. They have been modeled as an imaginary term in the speed of sound of the domain. The black dashed line indicates the $Q_{tot}$ measured in the experiment. The intersection between both lines leads us to estimate an air loss factor $\eta = 3.9\cdot 10^{-3}$ in our system. Controlling the environmental conditions (humidity and temperature) \cite{harris1966absorption} and the material's quality, the $Q$ factor might be further improved, enhancing the sensitivity of the device. Crucially, the measured $Q=128$ is achieved for a BIC in a fully 3D open acoustic environment, where the field couples to the continuum in all directions; it therefore should not be directly compared with values from 1D or 2D platforms, whose confinement suppresses radiation along one or more directions. Nonlinear far-field excitation accesses the symmetry-protected mode without structural symmetry breaking or the additional radiative leakage of a quasi-BIC. The significance of this $Q$ thus lies not in its numerical value alone, but in achieving resonant enhancement while preserving true-BIC radiation protection in a genuinely 3D open system.

\begin{figure}[h!]
    \makebox[\textwidth][c]{\includegraphics[width=5.5in,height=2.1in]{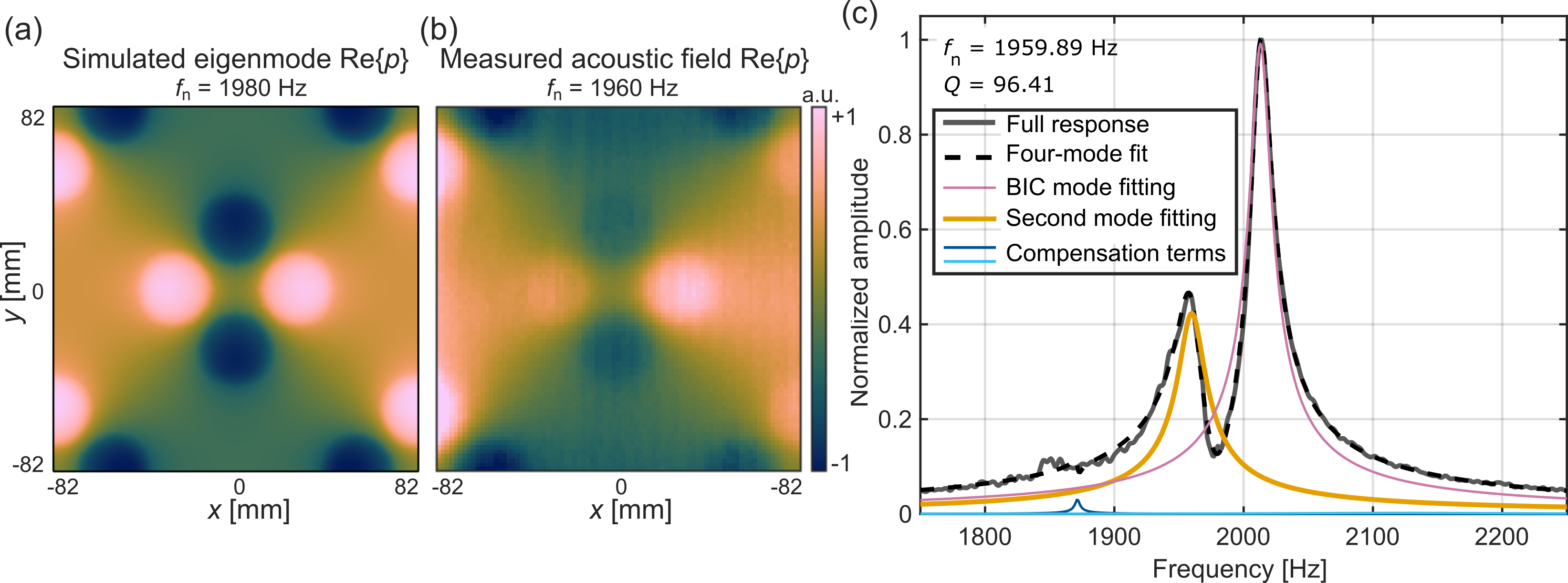}}
    \caption{\textbf{Second resonance characterization.} \textbf{a.} Simulated eigenmode for the experimental unit cell. \textbf{b.} Experimentally measured field. \textbf{c.} Curve fitting for estimating $Q$ factor properties of the second resonance.}
    \label{fig:figure_4_main_article}
\end{figure}

In addition to the primary mode at $2.103\ \text{kHz}$, a second mode is observed near $1.959\ \text{kHz}$ in Fig. \ref{fig:figure_2_main_article}(d). The corresponding field distribution for this mode is shown in Fig. \ref{fig:figure_4_main_article}(b). Unlike the BIC mode, this mode exhibits stronger relative field intensity near the semi-complete holes along the boundary rather than the central holes,  and possesses similar spatial symmetry. The existence of this resonance can be understood as a folded band originating from a regular bound state below the sound cone, consequence of the chosen rotated geometry for our experiment. More details can be found in Appendix \ref{secA5}. Despite not being a real BIC, this mode presents solid properties for being exploited in wave-related applications. The estimated $Q$ factor is $96.41$. 

\section{Conclusion}

Our results establish nonlinear frequency conversion as a general mechanism for accessing symmetry-protected states that are fundamentally inaccessible under linear excitation. More broadly, they show that nonlinear interactions can circumvent symmetry-imposed selection rules while preserving the protected nature of the underlying state. 
Our work opens a route towards exploiting BICs in fully open structures for wave-control applications, including sensing, energy harvesting and communications. By generating localized virtual sources through nonlinear ultrasonic demodulation, our approach enables far-field excitation without introducing additional radiative decay channels, thereby preserving the intrinsic non-radiative character and strong field enhancement of the BIC. This non-invasive excitation strategy leaves the resonator fully accessible to its surrounding environment, making it particularly attractive for sensing and other wave-control applications in which direct interaction with external perturbations is essential.

More generally, our results show that the symmetry protecting a bound state need not prevent its controlled excitation, provided that energy is generated locally rather than injected through the continuum. This concept establishes a general strategy for accessing symmetry-protected states while preserving their intrinsic properties.

\section{Acknowledgements}\label{secAcknowledgements}

We acknowledge Project No. CNS2023-145510 funded by MCIN/AEI/10.13039/501100011033, “European Union NextGenerationEU/PRTR”, Project No. CIPROM/2023/44 funded by Generalitat Valenciana, and Project No. PID2024-158832NB-C22. This work was supported by DYNAMO project (101046489), funded by the European Union. Views and opinions expressed are however those of the authors only and do not necessarily reflect those of the European Union or European Innovation Council. Neither the European Union nor the granting authority can be held responsible for them.


\vspace{1cm}
\noindent\rule{\linewidth}{0.4pt}
\vspace{1cm}

\begin{appendices}

\section{Theory}\label{secA1}
This appendix summarizes the mode-matching formulation used to compute the band structure and derives the symmetry conditions responsible for the existence of BICs under both normal and oblique incidence.

\subsection{Mode matching and band structure computation}

\begin{figure}[h!]
    \centering
    \includegraphics[width=0.5\linewidth]{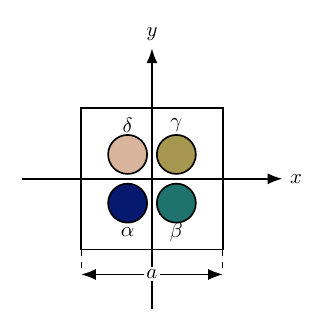}
    \caption{Geometry of the unit cell}
    \label{fig:unit_cell_supplementary}
\end{figure}

We consider an infinite rigid plate periodically patterned with identical unit cells, each containing four cylindrical boreholes arranged in a square lattice (Fig.~\ref{fig:unit_cell_supplementary}).
Applying Bloch's theorem, the pressure field as well as the normal derivative outside the cavity could be written as
\[
    \begin{aligned}
        \Psi_s (\mathbf{K}, \mathbf{r}, z) &= A_0 e^{-i k_0 z} e^{i \mathbf{K} \cdot \mathbf{r}} + \sum_{\mathbf{G} \neq \mathbf{0}} A_{\mathbf{G}} e^{-q_{\mathbf{G}} z} e^{i \mathbf{K}_{\mathbf{G}} \cdot \mathbf{r}}, \\
        v_n(\mathbf{K},\mathbf{r},z) &= -i \frac{1}{Z_b} A_0 e^{-i k_0 z} e^{i \mathbf{K} \cdot \mathbf{r}} - \sum_{\mathbf{G} \neq \mathbf{0}} \frac{q_{\mathbf{G}}}{k_0Z_b} A_{\mathbf{G}} e^{-q_{\mathbf{G}} z} e^{i \mathbf{K}_{\mathbf{G}} \cdot \mathbf{r}},
    \end{aligned}
\]

where $\mathbf{G} = \frac{2m\pi}{L_x}\hat{x} + \frac{2n\pi}{L_y}\hat{y}$ is the reciprocal lattice vector, $c_b$ is the velocity of sound, $\omega$ is the angular frequency, $k_0$ is the bulk wavenumber ($k_0 = \omega/c_b$) and $Z_b = c_b\rho_b$ is the acoustic bulk impedance. The variable $A_{\mathbf{G}}$ denotes the amplitude of the pressure field for each diffraction order, and $\mathbf{K}_{\mathbf{G}} = \mathbf{K} + \mathbf{G}$ is the transverse wave vector. The wavenumber in the $z$ direction $q_{\mathbf{G}}$  satisfies the dispersion relation $|\mathbf{K}_{\mathbf{G}}|^2 + q_{\mathbf{G}}^2 = k_0^2$. The first term represents the propagating zeroth diffraction order, whereas all higher diffraction orders are evanescent within the frequency range considered in the main text. \\
The pressure field inside the boreholes can be expanded in terms of cylindrical waveguide modes as
\[
    \Psi(r,z) = \sum_{q,n} B_{qn} J_q\left(\kappa_{qn}^{\prime}\frac{r}{R}\right) \cos[k_{z,qn}(z+L)],
\]
where $R$ is the radius of the boreholes, $\kappa_{qn}^{\prime}$ is the $n$-th root of the derivative of the Bessel function $J_q^{\prime}(\cdot)$, and $B_{qn}$ is the modal amplitude. Because the operating wavelength is much larger than the borehole radius, only the fundamental waveguide mode is retained. Consequently, the pressure and normal velocity inside the $l$-th borehole simplify to
\[
\begin{aligned}
    \Psi_l(z) &= B_l \cos[k_0(z+L)], \\
    V_{nl}(z) &= \frac{-B_l}{Z_b} \sin[k_0(z+L)].
\end{aligned}
\]
Each unit cell contains four identical boreholes, labeled $\alpha, \beta, \gamma,$ and $\delta$ in counter clockwise direction. We apply mode-matching technique as in \cite{torrent2012acoustic,torrent2018acoustic,marti2024observation}. Enforcing continuity of pressure and normal velocity at the interface and projecting the resulting equations onto the waveguide basis yields the secular system. 

\begin{align}     
\sum_{\mathbf{G}}A_Ge^{i(\mathbf{K+G})\cdot\mathbf{R}_\ell}H_{\ell}(\mathbf{K+G}) = B_\ell\cos{(k_0L)} &, \nonumber \\
    A_G = \frac{k_0}{q_G}\sum_\alpha f_\alpha H_\alpha^*(\mathbf{K+G})e^{-i(\mathbf{K+G})\cdot\mathbf{R}_\alpha}\sin{(k_0L)}B_\alpha, \nonumber
\end{align}

where $f$ denotes the filling fraction of the boreholes. The geometric form factor $H_l(\mathbf{K+G}) = \frac{1}{\Omega_\ell}\int e^{i(\mathbf{K+G})\cdot(\mathbf{r}-\mathbf{R}_\ell)}d\Omega_\ell$ can be further simplified considering all equal cylindrical boreholes with radius $R$
\[
    H_l(\mathbf{K+G}) = \frac{2 J_1(|\mathbf{K+G}| R)}{|\mathbf{K+G}| R},
\]
with $J_1(\cdot)$ the first-order Bessel function of the first kind.\\
By substituting the expression for $A_{\mathbf{G}}$ back into the first continuity equation to eliminate the external field coefficients, we can cast the secular equations into a compact matrix form $\mathcal{M} \mathbf{B} = \mathbf{0}$, where $\mathbf{B} = [B_\alpha, B_\beta, B_\gamma, B_\delta]^T$ is the amplitude vector. The matrix elements of $\mathcal{M}$ are given by 
\[
\mathcal{M}_{ij} = \delta_{ij} \cos(k_0 L) - f \chi_{ij} \sin(k_0 L).
\]
Here, $\chi_{ij}$ is the dimensionless coupling coefficient quantifying the radiative interaction between the $i$-th and $j$-th boreholes mediated by the external region, explicitly defined as

\[
\chi_{ij} = \sum_{\mathbf{G}}\frac{k_0}{q_G}H_i(\mathbf{K+G})H_j(\mathbf{K+G})e^{i(\mathbf{K+G})\cdot\mathbf{R}_{ij}}, \; i, j \in [\alpha,\beta,\gamma,\delta].
\]
\\

\subsection{BIC condition}

\begin{figure}[h!]
    \centering
    \includegraphics[width=0.7\linewidth]{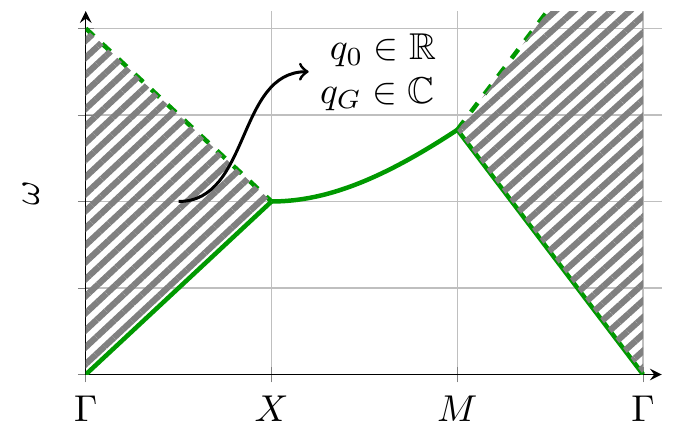}
    \caption{Square sound cone band structure with the first radiation zone indicated.}
    \label{fig:schematic_band_structure}
\end{figure}

Within the first radiation continuum (shaded region in Fig. (\ref{fig:schematic_band_structure})), only the zeroth diffraction order remains propagating, whereas all higher orders are evanescent. Consequently, a bound state in the continuum requires the complete suppression of the zeroth-order radiation amplitude $A_0$.

Given the $C_4$ rotational symmetry of the square unit cell, the coupling term $\chi_{ij}$ depends solely on the relative spatial separation between the boreholes ($\mathbf{R}_{ij}$). We can classify the interactions into self-coupling $\chi_{\alpha\alpha}$, nearest-neighbor coupling $\chi_{\alpha\beta}$, and diagonal coupling $\chi_{\alpha\gamma}$. This translation-invariant property dictates that the coupling matrix $\boldsymbol{\chi}$ is a symmetric circulant matrix, allowing us to express the full system $\mathcal{M} \mathbf{B} = \mathbf{0}$ as:
\[
    \left( \cos(k_0 L) \mathbf{I} - f \sin(k_0 L) 
    \begin{bmatrix}
        \chi_{\alpha\alpha} & \chi_{\alpha\beta} & \chi_{\alpha\gamma} & \chi_{\alpha\beta} \\
        \chi_{\alpha\beta} & \chi_{\alpha\alpha} & \chi_{\alpha\beta} & \chi_{\alpha\gamma} \\
        \chi_{\alpha\gamma} & \chi_{\alpha\beta} & \chi_{\alpha\alpha} & \chi_{\alpha\beta} \\
        \chi_{\alpha\beta} & \chi_{\alpha\gamma} & \chi_{\alpha\beta} & \chi_{\alpha\alpha} 
    \end{bmatrix} \right)
    \begin{bmatrix} 
        B_\alpha \\ B_\beta \\ B_\gamma \\ B_\delta 
    \end{bmatrix} 
    = \mathbf{0},
\]
where $\mathbf{I}$ is the $4 \times 4$ identity matrix. By virtue of the properties of circulant matrices, the eigenvectors are analytically known as discrete Fourier modes:
\begin{equation}
    B_n = B_0 e^{i 2\pi n \ell / N}, \quad n, \ell \in \{0, 1, 2, 3\},
    \label{eq:ansatz}
\end{equation}
where $N=4$ is the total number of boreholes in the unit cell, $n$ is the spatial index of the borehole, and $\ell$ is the azimuthal mode index determining the phase symmetry of the collective resonance. 

Thus, the corresponding decoupled dispersion relations for each symmetry mode $\ell$ evaluate to:
\begin{equation}
    \cos(k_0 L) - f \sin(k_0 L) \sum_{n=0}^{3} \chi_{0,n} e^{i 2\pi n \ell / 4} = 0,
    \label{eq:secular_equation_four_fold}
\end{equation}
where $\chi_{0,n}$ denotes the first row of the coupling matrix $\boldsymbol{\chi}$.\\

Let us start by considering normal incidence to our phononic crystal ($\mathbf{K} = \mathbf{0}$, i.e., $\Gamma$ point in the band structure). The expression for the radiation amplitude $A_0$ can be rewritten as 

\[
    A_0 = f \sin{(k_0L)}\sum_\alpha B_\alpha. 
\]

Thus, the condition for having a BIC at the $\Gamma$ point is equivalent to $\sum_\alpha B_\alpha = 0$. Considering the ansatz in equation (\ref{eq:ansatz}), we can clearly see that except for the fundamental mode $\ell = 0$, the net monopole radiation into the external plane wave ($A_0$) cancels out due to phase mismatch. 

The eigenfrequencies for the modes can be computed using equation (\ref{eq:secular_equation_four_fold}) with $\mathbf{K} = \mathbf{0}$. The terms $\chi_{0,n}$ become lattice sums

\[
    \chi_{0,n} = \sum_{\mathbf{G}}\frac{k_0}{q_G}H_0(\mathbf{G})H_n(\mathbf{G})e^{i\mathbf{G}\cdot\mathbf{R}_{0n}}, \; n \in \{0,3\}
\]

It is worth noting that the preceding discussion assumed a normally incident wave $\mathbf{K}=\mathbf{0}$. However, symmetry-protected BICs can still survive under oblique incidence $\mathbf{K} \neq \mathbf{0}$, provided that the incident plane wave is orthogonal to a specific symmetry axis of the unit cell. 
To intuitively understand this phenomenon, consider first a simplified unit cell containing only two identical boreholes aligned along the $x$-axis. If the incident wave is obliquely launched strictly along the $y$-axis $\mathbf{K}=K_y\mathbf{\hat{y}}$, the macroscopic zeroth-order radiation amplitude $A_0$ becomes a phase-modulated superposition:
\[
    A_0 = f \sin{(k_0L)}\frac{2J_1(|\mathbf{K}|R)}{|\mathbf{K}|R}\sum_{\ell \in \{1,2\}} B_\ell e^{-i\mathbf{K}\cdot\mathbf{R}_\ell}
\]
Because the wave vector $\mathbf{K}$ is perpendicular to the alignment axis of the boreholes ($\mathbf{K} \cdot \mathbf{R}_{1,2} = 0$), no spatial phase difference accumulates between them. Consequently, exciting the asymmetric mode strictly cancels the net radiation $A_0 = 0$.\\ 
This physical mechanism can be naturally extended back to our four-fold symmetry system ($\alpha, \beta, \gamma, \delta$). Although an arbitrary oblique wave vector would break the global $C_4$ rotational symmetry and inevitably induce radiative leakage, aligning $\mathbf{K}$ along a direction that preserves a mirror symmetry orthogonal to the propagation direction. In the case of the $C_4$ rotational symmetry system, it will happen in three different directions of the dispersion manifold: $\mathbf{K} = K_x\hat{x}$, $\mathbf{K} = K_y\hat{y}$ and $\mathbf{K} = K(\hat{x}+\hat{y})$, with the first and the third belonging to the IBZ path, namely $\Gamma-X$ and $M-\Gamma$ directions. 

Without loss of generality, we consider the case $\mathbf{K} = K_x\hat{x}$, as the reasoning for the other two cases is analogous. In this case, boreholes located on the same horizontal wavefront namely, the pair $\alpha$ and $\delta$, and the pair $\beta$ and $\gamma$ share identical spatial phases. By exciting a specific asymmetric mode where the pressure amplitudes within these mirror-symmetric pairs are strictly out-of-phase, the pairwise radiation exactly cancels out:
\[
\begin{aligned}
    &B_\alpha e^{-i K_x x_\alpha} + B_\delta e^{-i K_x x_\delta} = e^{-i K_x x_\alpha}(B_\alpha + B_\delta) = 0, \\
    &B_\beta e^{-i K_x x_\beta} + B_\delta e^{-i K_x x_\gamma} = e^{-i K_x x_\beta}(B_\beta + B_\gamma) = 0.
\end{aligned}
\]
Thus, the zero net monopole radiation condition is robustly preserved for the entire unit cell. This pairwise cancellation mechanism ensures that the symmetry-protected BIC survives as a dispersion band even under oblique incidence, provided the wave vector is restricted to the mirror-symmetric axes. The out-of-phase relation arises as a consequence of the system's mirror symmetry. Because the wave vector lies along a mirror-symmetric direction, the mirror operator commutes with the secular matrix. Therefore the eigenstates have definite mirror parity and they can be decoupled into two independent orthogonal subspaces: the symmetric subspace and the antisymmetric subspace.\\
Under this pairwise asymmetric condition, the fully coupled $4 \times 4$ secular equation $\mathcal{M}(\mathbf{K})\mathbf{B} = \mathbf{0}$ elegantly decouples. The equations for $B_\delta$ and $B_\gamma$ become linearly dependent on those for $B_\alpha$ and $B_\beta$, rigorously reducing the $4 \times 4$ system to a $2 \times 2$ subspace:
\[
    \begin{bmatrix}
        \cos(k_0 L) - f \Delta\chi_v \sin(k_0 L) & - f \Delta\chi_h \sin(k_0 L) \\
        - f \Delta\chi_h^* \sin(k_0 L) & \cos(k_0 L) - f \Delta\chi_v \sin(k_0 L)
    \end{bmatrix}
    \begin{bmatrix}
        B_\alpha \\ B_\beta
    \end{bmatrix}
    = \mathbf{0},
\]
where we have $\Delta\chi_v=\chi_{\alpha\alpha}-\chi_{\alpha\delta}$ and $\Delta\chi_h=\chi_{\alpha\beta}-\chi_{\alpha\gamma}$. Due to the mirror symmetry across the $y$-axis, the cross-coupling coefficients satisfy $\chi_{\beta\alpha} - \chi_{\gamma\alpha} = (\chi_{\alpha\beta} - \chi_{\alpha\gamma})^*$. Furthermore, given the geometric symmetry along the $x$-axis between the top and bottom pairs, the diagonal elements are identical ($\chi_{\gamma\gamma} = \chi_{\alpha\alpha}$ and $\chi_{\gamma\delta} = \chi_{\alpha\beta}$). Setting the determinant of this Hermitian reduced matrix to zero yields two completely uncoupled, real-valued analytical dispersion relations:
\begin{equation}
    \cos(k_0 L) - f \sin(k_0 L) \left[ \Delta\chi_v \pm |\Delta\chi_h| \right] = 0.
    \label{eq:secular_BIC_equation}
\end{equation}
Here, the $\pm$ sign corresponds to the symmetric and asymmetric modes between the left and right pairs. Equation (\ref{eq:secular_BIC_equation}) explicitly demonstrates that the dispersion branches associated with the symmetry-protected BIC remain decoupled from the radiation continuum as long as the mirror symmetry of the excitation direction is preserved.\\


\section{Ultrasonic Modulation and Demodulation}\label{secA2}
This appendix establishes that nonlinear demodulation of an amplitude-modulated ultrasonic field generates an audio-frequency component that preserves the phase integrity of the original modulation signal. This property underpins the non-invasive excitation scheme employed throughout the manuscript. 

The basic working principle of the mixer-parametric speaker system is to modulate the audible sound with an ultrasound carrier component. Then, by leveraging the nonlinearity of local high-intensity acoustic field, the targeting low-frequency component will be demodulated at certain impact surface or region (i.e. the BIC holes in the actual design). Specifically, two important aspects of the ultrasound modulation-demodulation process will be validated: whether demodulation can indeed regenerate the low-frequency component and whether the demodulated signal is in phase with the original signal.

The actual input audio signal can be modeled as a tonal frequency $f_s$ that is modulated by an enveloped signal:
\begin{equation*}
    s(t) = g(t) \cos(\Omega t+\phi_s).
\end{equation*}
In the case of our experiment, $g(t)$ is a Gaussian-envelope pulse (i.e. real-valued positive envelope), $\Omega = 2\pi f_s = 2\pi \times 2\ \text{kHz}$ is the angular tonal frequency of the input signal, and $\phi_s$ is the controlled phase. As $g(t)$ does not affect any frequency and phase analysis, without loss of generality, it will be set to $1$ for simplicity in the following discussion.

The audio frequency signal is then mixed (i.e. amplitude modulated) with the ultrasound carrier wave with an angular frequency of $\omega = 2\pi \times 40\ \text{kHz}$ and carrier phase of $\phi_c$. Because the carrier phase is not controlled experimentally, it is treated as an arbitrary constant.

To illustrate the demodulation mechanism, we consider an audio-frequency-modulated one-dimensional plane wave ultrasound field including both forward- and  backward-propagating components:
\begin{align*}
    p(x,t) &= \left[1+m\cos\left(\Omega t+\phi_s\right)\right]\left[\cos\left(\omega t+\phi_{cf}-kx \right)+\cos\left(\omega t+\phi_{cb}+kx \right)\right]
\end{align*}
where $m<1$ is the modulation depth and $\phi_{ci}$ for $i\in\{f,b\}$ is the carrier phase for the forward- and backward-propagating components respectively. Expanding the expression above, we obtain six terms: the first pair $T_{1i}$ corresponds to the carrier wave components, the second pair $T_{2i}$ corresponds to the upper sideband ($\varpi^+=\omega+\Omega$), and the third pair $T_{3i}$ corresponds to the lower sideband ($\varpi^-=\omega-\Omega$):
\begin{align*}
    p(x,t) &= \cos\left(\omega t+\phi_{cf}-kx \right) + \cos\left(\omega t+\phi_{cb}+kx \right)\\
    &\quad +\frac{m}{2}\cos(\varpi^+t+\theta_f^+ -kx) + \frac{m}{2}\cos(\varpi^+t+\theta_b^+ +kx)\\ 
    &\quad +\frac{m}{2}\cos(\varpi^-t+\theta_f^- -kx) + \frac{m}{2}\cos(\varpi^-t+\theta_b^- +kx)\\
    &\equiv T_{1f} + T_{1b} + T_{2f} + T_{2b} + T_{3f} + T_{3b},\\
    &= \sum_{i\in\{f,b\}}(T_{1i}+T_{2i}+T_{3i})
\end{align*}
where $\theta_i^\pm = \phi_{ci}\pm\phi_s$ for $i \in \{f,b\}$.

In an arbitrary high-acoustic-intensity environment, the linear acoustic wave equation will fail to adequately describe the pressure field. However, the sound behavior can be well-approximated by the Westervelt equation:
\begin{equation*}
    \nabla^2p - \frac{1}{c^2} \frac{\partial^2p}{\partial t^2} = -\frac{\beta}{\rho_0c_0^4} \frac{\partial^2\left(p^2\right)}{\partial t^2},
\end{equation*}
where $\beta$ is the second-order nonlinearity coefficient. The term on the right-hand side represents a source term, which corresponds to the generation of new frequency components. Expressing the squared pressure field $p^2$ following the previous expansion, we obtain a series of terms. The only terms relevant to the present discussion are those oscillating at the modulation frequency $\Omega$ ($2T_{1i}T_{2j}$ and $2T_{1i}T_{j3}$), since these correspond to the regenerated audible signal:

\begin{align*}
    T_{1i}^2 &= \frac{1}{2} + \frac{1}{2}\cos[2\omega t + 2\phi_{ci} + 2\text{sgn}(i)kx]\\
    T_{2i}^2 &= \frac{m^2}{8} + \frac{m^2}{8}\cos[2\varpi^+t + 2\theta_i^+ +2\text{sgn}(i)kx]\\
    T_{3i}^2 &= \frac{m^2}{8} + \frac{m^2}{8}\cos[2\varpi^-t + 2\theta_i^-+2\text{sgn}(i)kx]\\
    2T_{1i}T_{2j} &= \frac{m}{2}\cos[(2\omega+\Omega)t + \phi_{ci} + \theta_j^+ +(\text{sgn}(i)+\text{sgn}(j))kx]\\
    &\quad + \frac{m}{2}\cos[\Omega t + \phi_s + (\text{sgn}(i)+\text{sgn}(j))kx]\\
    2T_{1i}T_{3j} &= \frac{m}{2}\cos[(2\omega-\Omega)t + \phi_{ci} + \theta_j^- +(\text{sgn}(i)+\text{sgn}(j))kx]\\
    &\quad + \frac{m}{2}\cos[\Omega t + \phi_s + (\text{sgn}(i)+\text{sgn}(j))kx]\\
    2T_{2i}T_{3j} &= \frac{m^2}{4}\cos[2\omega t + \theta_i^+ + \theta_j^- +(\text{sgn}(i)+\text{sgn}(j))kx]\\
    &\quad + \frac{m^2}{4}\cos[2\Omega t+2\phi_s+(\text{sgn}(i)+\text{sgn}(j))kx]\\
    2T_{1f}T_{1b} &= \cos(2\omega t + \phi_{cf} + \phi_{cb} + \cos(\phi_{cf} - \phi_{cb} - 2kx)\\
    2T_{2f}T_{2b} &= \frac{m^2}{4}\cos(2\varpi^+ t + \theta_f^+ + \theta_b^+) + \frac{m^2}{4}\cos(\theta_f^+ - \theta_b^+ - 2kx)\\
    2T_{3f}T_{3b} &= \frac{m^2}{4}\cos(2\varpi^- t + \theta_f^- + \theta_b^-) + \frac{m^2}{4}\cos(\theta_f^- - \theta_b^- - 2kx)
\end{align*}
For compactness, we have defined

\begin{align*}
    \text{sgn}(i) = \begin{cases}
        -1, & \text{ for } i=f \\
        +1, & \text{ for } i=b
    \end{cases}
\end{align*}

This is the demodulation process where the audio signal, originally encoded in the amplitude of the ultrasonic carrier, is recovered as an independent acoustic wave through the nonlinear interactions. The demodulated component takes the form of $p_{\text{audio}}\propto \frac{m}{2}\cos[\Omega t + \phi_s + (\text{sgn}(i)+\text{sgn}(j))kx]$, where $(\text{sgn}(i)+\text{sgn}(j))kx$ only introduces a spatial phase shift depending on propagation direction. Crucially, the recovered audio phase is always of $\phi_s$, independent of both carrier phases $\phi_{cf}$ and $\phi_{cb}$. This confirms that the demodulated pulse observed at the BIC aperture is phase-aligned with the originally transmitted signal, regardless of the random carrier phase introduced by the mixer.

\section{Characterization of the Focused Parametric Speaker Domes}\label{secA3}
While the plane wave model as outlined in Appendix \ref{secA2} establishes the theoretical basis for the modulation-demodulation process, efficient demodulation in practice requires sufficiently high acoustic intensity to provide strong nonlinear interactions at the target spot. Moreover, to concentrate the ultrasonic energy at the BIC holes while minimizing nonlinear interactions elsewhere, the parametric speaker array was designed on a spherical dome geometry, with all transducer elements oriented toward a common focal point at the center of curvature. Each dome has a nominal focal distance of $195\ \text{mm}$ and aperture diameter of $53.5\ \text{mm}$, subtending an aperture solid angle of $30^\circ$. One spherical substrate accommodates $60$ counts of $40\ \text{kHz}$ transducers arranged in 4 concentric layers (Fig.C3a), and four identical domes were fabricated for the full BIC experiment.

\begin{figure}[h!]
    \makebox[\textwidth][c]{\includegraphics[width=5.5in,height=5.13in]{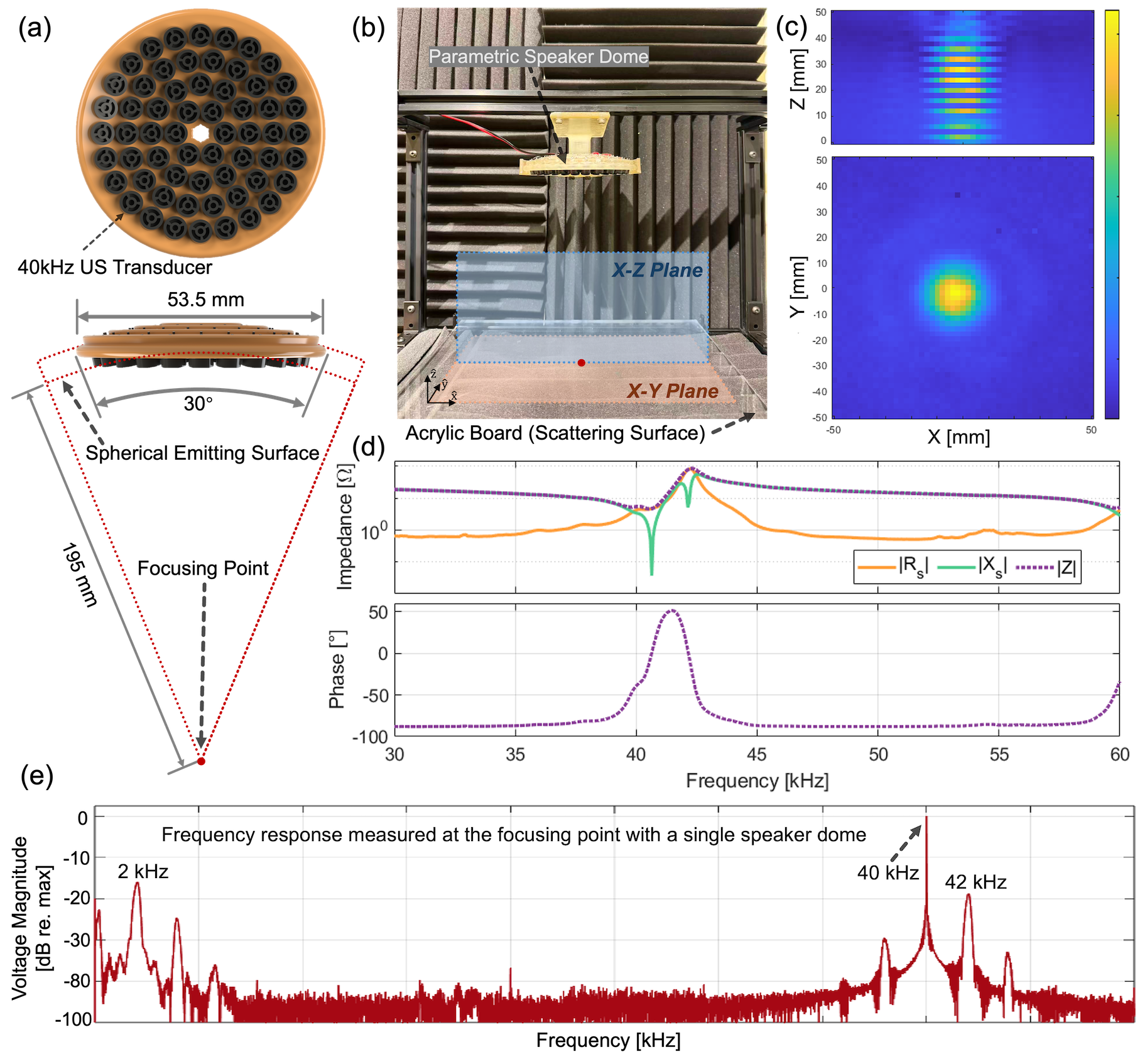}}
    \caption{Dome Characterization}
    \label{fig:Dome Characterization}
\end{figure}

To characterize the generated focused field, a single dome was oriented downward toward an acrylic scattering board placed at the nominal focal distance of $195\ \text{mm}$ (Fig.C3b). The demodulated audio field was scanned in both the X-Y (top) and X-Z (frontal) planes over areas of $100\times100\ \text{mm}^2$ and $100\times50\ \text{mm}^2$, respectively, using the same microphone as in the full BIC field scan, capped with a straight aluminum needle tube (inner diameter of $1.70\ \text{mm}$). At each grid point, a Gaussian pulse centered at $2\ \text{kHz}$ with an envelope width of $7.5\ \text{ms}$ and pulse length of $200\ \text{ms}$ was transmitted. Each received signal was individually filtered using a zero-phase notch filter at $40\ \text{kHz}$ and a zero-phase lowpass filter with a $5\ \text{kHz}$ cutoff were applied to isolate the demodulated audio component. This process was repeated $12$ times on a single point to get an average signal for random noise reduction purpose.

The measured field map confirms that the focused energy is well-concentrated at the designated focusing point, $195\ \text{mm}$ away from the emitting surface, forming an approximately oval-shaped parcel with a cross-sectional diameter of roughly $10\ \text{mm}$ (Fig.C3c). Since the designed BIC hole has a diameter of approximately $30\ \text{mm}$, the focal spot can fit comfortably within the hole by precisely positioning each dome at the appropriate tilt angle and height, ensuring minimal acoustic leakage onto the surrounding BIC surface.

The electrical impedance of the fabricated parametric speaker dome is shown in Fig.C3d. The measured resistance, reactance, and impedance magnitude exhibit a pronounced electromechanical resonance near the $40\ \text{kHz}$ operating frequency, confirming that the assembled dome remains strongly responsive within the intended ultrasonic excitation band.

Figure C3e shows the frequency-domain voltage response measured at the focal point using a single speaker dome. The spectrum contains a dominant component near the $40\ \text{kHz}$ carrier frequency, while the demodulated audio component near $2\ \text{kHz}$ is approximately $15.6\ \text{dB}$ weaker. This relatively low audio-frequency amplitude indicates that the desired demodulated signal is substantially weaker than the residual ultrasonic carrier and associated high-frequency components. Therefore, appropriate signal filtering is required to suppress the carrier component and reliably extract the audio-frequency response used in the subsequent BIC field measurements.

\section{Experimental Setup And Signal Processing}\label{secA4}
\subsection{BIC Metamirror Fabrication}
For ease of fabrication, the designed BIC metamirror was truncated at the boundary of its second Brillouin zone. The structure was fabricated using an ELEGOO Saturn 3 3D printer with standard photopolymer resin, which has a Young's modulus of $708\ \text{MPa}$ and a Shore hardness of $88\text{D}$ after full curing. The truncated edges were treated as hard boundaries, approximating an infinitely extended metamirror under the experimental conditions.

\subsection{Gaussian Pulse Definition and Filtering Scheme}
The excitation signal was defined as a Gaussian pulse. At each measurement point, 20 identical regularized Gaussian pulses centered at $2\ \text{kHz}$ were transmitted through the parametric speaker array. Each pulse had an envelope width of $7.5\ \text{ms}$, a total duration of $200\ \text{ms}$, and an onset time of $20\ \text{ms}$.

The received signals were processed using a two-stage filtering scheme. First, a zero-phase notch filter centered at $40\ \text{kHz}$ with a bandwidth of $0.75\ \text{kHz}$ was applied to suppress the ultrasonic carrier component. A low-pass filter with a cutoff frequency of $5\ \text{kHz}$ was then used to retain the audio-frequency response. The 20 filtered signals were subsequently averaged to obtain a representative field response at each scanning point.

\subsection{Signal Pathway}
The excitation signal defined above was generated in MATLAB and output through a data acquisition card (DAQ, NI-6353). It was then routed to two $40\ \text{kHz}$ ultrasonic mixer amplifiers (Audfly), each driving two identical parametric speaker domes. To reproduce the alternating-phase distribution of the designed BIC eigenmode, the two mixers received oppositely phased signals from the DAQ. Consequently, channels 1 and 3 generated positively phased modulated ultrasound, whereas channels 2 and 4 generated negatively phased modulated ultrasound. The detailed modulation model is provided in Appendix \ref{secA2}. The resulting modulated ultrasonic signals were used to excite the acoustic field above the metamirror.

The acoustic field was measured using a microphone (PCB Piezotronics 378C01) mounted on a three-axis motorized stage. The microphone was fitted with a customized thin aluminum tube with an inner diameter of $1.70\ \text{mm}$ and a $90^\circ$ bend at the tip to minimize disturbance of the acoustic field by the overhead speaker arrays. The measured signal was conditioned using a PCB Piezotronics signal conditioner and returned to the DAQ for acquisition.

\subsection{2D Scanning Route}
For full-field measurements, the microphone was moved across the upper surface of the BIC metamirror along a zigzag scanning path with a step size of $2\ \text{mm}$. The scan covered an area of $162\times162\ \text{mm}^2$, corresponding to an $83\times83$ measurement grid. To efficiently capture the confined acoustic field while avoiding contact between the scanning probe and the metamirror, the microphone tip was maintained at a vertical distance of $2\ \text{mm}$ above the surface.

\subsection{Signal Time Gating}
Since the BIC field is excited by four parametric speaker domes focusing into the central holes, a hard time-gating was applied to the measured signal to eliminate contributions from the incident field and so to isolate the scattered BIC response. Specifically, the received signal at each scanning point was truncated to retain only the portion after the incident field had decayed to $1.5\%$ of its peak value. Because the incident and scattered fields can vigrosely mix inside the BIC central holes and cannot be easily separated, the incident field was approximated by measuring the response of a single dome sending the same Gaussian pulse to a planar scattering board while following the same general setup. With this time-gating method, all received signals were truncated beyond a critical time of $t = 0.0310\ \text{s}$.

\subsection{Curve-fitting by Lorentzian Resonance}
To quantify the quality factor (Q-factor) of the observed BIC modes, a curve-fitting procedure was performed in the frequency domain. The scattered field spectrum was modeled as a superposition of four Lorentzian resonances, each with a frequency-domain representation of:
\begin{equation}
    S(\omega) = \frac{B\exp(j\phi)}{j(\omega-\omega_r)+\omega_i},
\end{equation}
where $B$ and $\phi$ are the signal amplitudes and phase, $f_r$ is the resonance frequency, and $f_i$ is the half-linewidth. Out of the four fitting signals, two of Lorentzians target the high-amplitude modeling (corresponding to 2 BIC modes), while the remaining two serve as compensation terms to account for background contributions. The fitting was performed using MATLAB's nonlinear least-square optimization, and the resulting parameters are summarized in Table 1:
\begin{table}[h]
\centering
\caption{Optimized Lorentzian Fitting Parameters}
\begin{tabular}{cccccc}
\hline
Peak Number & $B$ & $\phi$ [rad] & $f_r = 2\pi\omega_r$ [Hz] & $f_i = 2\pi\omega_i$ [Hz] & $Q$ \\
\hline
1 & $\sim$0.0000 & $-0.2763$ & 1871.42 & 1.66 & 563.44 \\
2 & 0.0015 & $-0.0199$ & 1959.89 & 10.16 & 96.41 \\
3 & 0.0026 & $0.1001$ & 2013.02 & 7.85 & 128.24 \\
4 & 0.0001 & $-0.8185$ & 2120.65 & 81.78 & 12.97 \\
\hline
\end{tabular}
\end{table}\\

\section{Second Resonance Apparition}\label{secA5}
This appendix explians the origin of the second resonance observed experimentally. We show that it is not an additional bound state in the continuum, but rather a consequence of the rotated experimental geometry. Specifically, the $45^{\circ}$ rotation and associated enlargement of the primitive cell induce a band-folding mechanism that maps a conventional bound state onto the $\Gamma$ point of the experimental Brillouin zone.

\begin{figure}[h!]
    \makebox[\textwidth][c]{\includegraphics[width=0.9\textwidth]{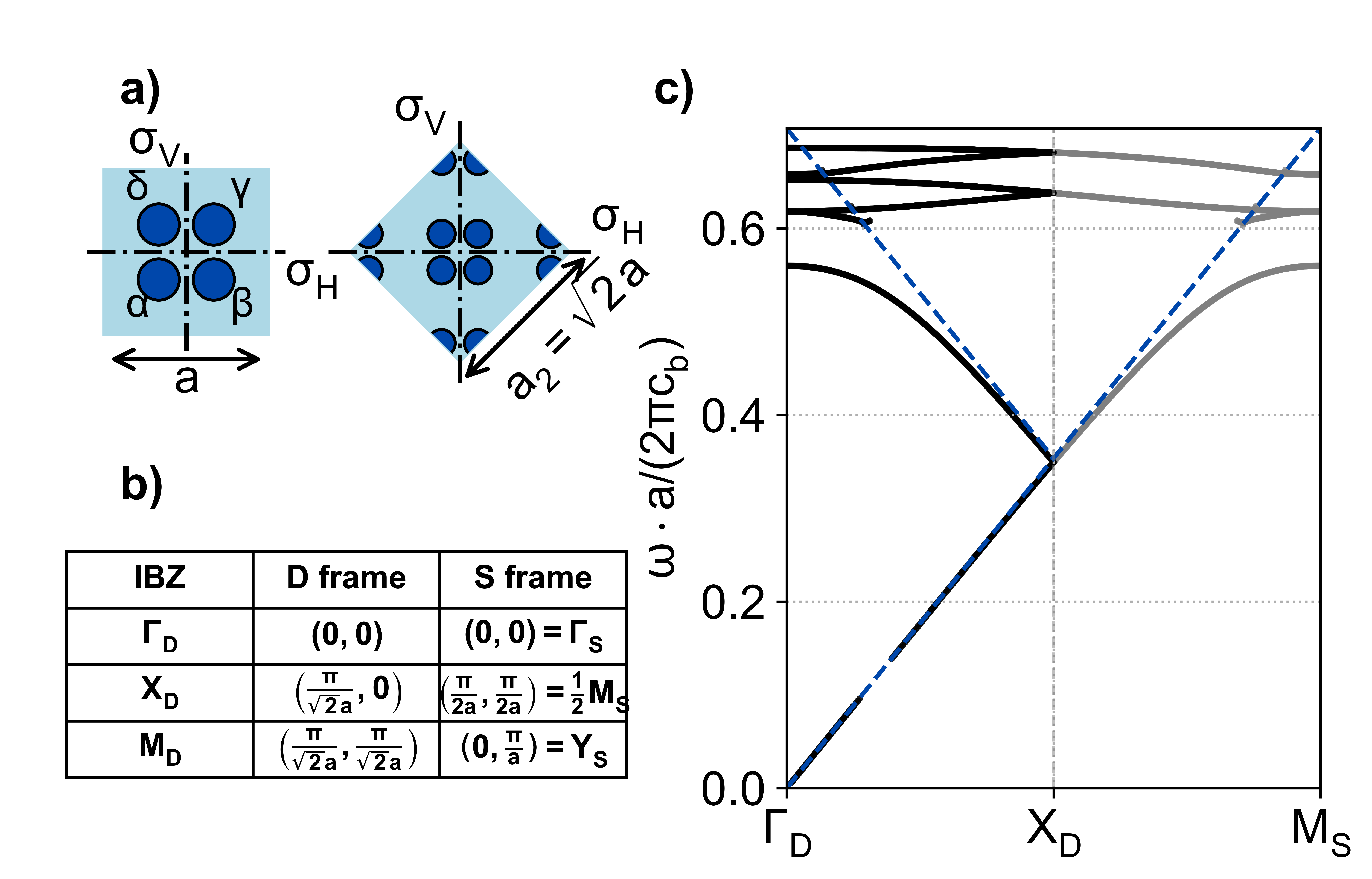}}
    \caption{Band folding mechanism for the apparition of the second peak. Panel a) shows the unit cell for both the theory (left) and the experiment (right). Table b) indicates the position in reciprocal space of the high-symmetry points of the experimental unit cell (D frame), and their coordinates in the theory frame (S frame), relating the points to those of the original IBZ. Panel c) represents the first part of the band structure for the experimental unit cell ($\Gamma_D - X_D$) and relates it to the direction $\Gamma_S - M_S$ via a band folding mechanism.}
    \label{fig:supplementary_4_figure_1}
\end{figure}

To analyse this effect, we consider a periodic lattice whose unit cell coincides with that employed in the experimental implementation (Fig. \ref{fig:supplementary_4_figure_1}(a) right unit cell). Our new unit cell is obtained after a rotation of $45^{\circ}$ and a scaling of $\sqrt{2}$ of the original lattice vector. For clarity, we refer to the reciprocal bases associated with the original square lattice and the rotated experimental lattice as the $S$ and $D$ frames, respectively. The lattice vectors of the D frame are related to those of the S frame through

\[
T = \begin{bmatrix}
    1 & 1 \\
    -1 & 1
\end{bmatrix}.
\]

The enlarged primitive cell defines a reduced Brillouin zone. Expressing the high-symmetry points of this Brillouin zone in the reciprocal basis of the original lattice reveals a direct correspondence between the two descriptions (Fig. \ref{fig:supplementary_4_figure_1}(b)). 

Consequently, the reciprocal-space path $\Gamma_D - X_D$ spans only half the reciprocal distance covered by the original $\Gamma_S - M_S$ path. The remaining half is therefore folded back into the reduced Brillouin zone.

As a consequence of this folding, the modes originally located at the $M_S$ point of the square lattice are mapped onto the $\Gamma_D$ point of the experimental Brillouin zone.  Fig.\ref{fig:supplementary_4_figure_1}(c) illustrates how the upper branch of the original dispersion relation folds onto the reduced Brillouin zone, giving rise to a second resonance with the same spatial symmetry as the designed BIC. Unlike the symmetry-protected BIC, however, this additional resonance originates from a conventional bound state below the sound cone and therefore does not constitute an independent BIC. Its predicted frequency lies below that of the target BIC, in agreement with the experimental observations. 

Because the theoretical analysis in the main text employs the primitive square unit cell, this folded branch does not appear in the corresponding band diagram. It emerges only when the enlarged experimental unit cell is adopted, demonstrating that the additional resonance is a geometrical consequence of the experimental implementation rather than a distinct physical mechanism.

\end{appendices}

\end{document}